\documentclass[10pt,conference]{IEEEtran}
\IEEEoverridecommandlockouts

\usepackage{cite}
\usepackage{amsmath,amssymb,amsfonts}
\usepackage{algorithmic}
\usepackage{graphicx}
\usepackage{textcomp}
\usepackage{xcolor}

\usepackage{xspace}
\ifCLASSOPTIONcompsoc
\usepackage[caption=false,font=normalsize,labelfont=sf,textfont=sf]{subfig}
\else
\usepackage[caption=false,font=footnotesize]{subfig}
\fi
\usepackage{colortbl}
\usepackage{multirow}
\usepackage{booktabs}
\usepackage{siunitx}
\usepackage{enumitem}
\usepackage{soul}

\usepackage{tikz}

\newcommand{\minisection}[1]{\vspace{1ex}\noindent {\bf #1}}

\usepackage[most]{tcolorbox}
\usepackage{etoolbox} %

\usepackage{hyperref}

\usepackage{fontawesome}

\definecolor{chopper_ins_color}{RGB}{255, 247, 240}
\newcounter{insightcounter}
\newcommand{\insight}[1]{
  \stepcounter{insightcounter}
  \begin{tcolorbox}[
    colframe=gray,
    colback=chopper_ins_color,
    boxrule=0.5pt,
    arc=3pt,
    left=0pt,
    right=0pt,
    top=0pt,
    bottom=0pt
  ]
  \textbf{\bf\faLightbulbO\ Insight \arabic{insightcounter}}: #1
  \end{tcolorbox}
}

\definecolor{chopper_rec_color}{RGB}{240, 247, 255}
\newcounter{reccounter}
\newcommand{\rec}[1]{
  \stepcounter{reccounter}
  \begin{tcolorbox}[
    colframe=gray,
    colback=chopper_rec_color,
    boxrule=0.5pt,
    arc=3pt,
    left=0pt,
    right=0pt,
    top=0pt,
    bottom=0pt
  ]
  \textbf{\faWrench\ Rec. \arabic{reccounter}}: #1
  \end{tcolorbox}
}

\definecolor{chopper_obs_color}{RGB}{247, 247, 247}
\newcounter{observationcounter}
\newcommand{\observation}[1]{
  \refstepcounter{observationcounter}
  \begin{tcolorbox}[
    colframe=gray,
    colback=chopper_obs_color,
    boxrule=0.5pt,
    arc=3pt,
    left=0pt,
    right=0pt,
    top=0pt,
    bottom=0pt
  ]
  \textbf{\faSearch\ Observation \arabic{observationcounter}}: #1
  \end{tcolorbox}
}

\usepackage{lipsum}

\definecolor{red}{RGB}{255, 102, 102}
\definecolor{orange}{RGB}{255, 178, 102}
\definecolor{yellow}{RGB}{255, 255, 153}
\definecolor{green}{RGB}{204, 255, 204}

\definecolor{copy}{HTML}{7570b3}
\definecolor{tech}{HTML}{d95f02}
\definecolor{market}{HTML}{1b9e77}

\usepackage[linecolor=red,backgroundcolor=red!25,bordercolor=red,textsize=tiny]{todonotes}

\newcommand{\name}{\textit{Chopper}\xspace}
\newcommand{\thetitle}{\name: A Multi-Level GPU Characterization Tool \& Derived Insights Into LLM Training Inefficiency}

\newcommand{\shaizeenST}[1]{{\st{#1}}} 
\renewcommand{\shaizeenST}[1] {}

\begin{document}

\title{\thetitle}

\author{
    \IEEEauthorblockN{Marco Kurzynski}
    \IEEEauthorblockA{
    \textit{University of Central Florida}\\
    marco.kurzynski@ucf.edu}

    \and

    \IEEEauthorblockN{Shaizeen Aga}
    \IEEEauthorblockA{
    \textit{Advanced Micro Devices, Inc.}\\
    shaizeen.aga@amd.com}

    \and

    \IEEEauthorblockN{Di Wu}
    \IEEEauthorblockA{
    \textit{University of Central Florida}\\
    di.wu@ucf.edu}
}

\maketitle

\thispagestyle{plain}
\pagestyle{plain}

\begin{abstract}

Training large language models (LLMs) efficiently requires a deep understanding of how modern GPU systems behave under real-world distributed training workloads. While prior work has focused primarily on kernel-level performance or single-GPU microbenchmarks, the complex interaction between communication, computation, memory behavior, and power management in multi-GPU LLM training remains poorly characterized.
In this work, we introduce \name{}, a profiling and analysis framework that collects, aligns, and visualizes GPU kernel traces and hardware performance counters across multiple granularities (i.e., from individual kernels to operations, layers, phases, iterations, and GPUs). Using \name{}, we perform a comprehensive end-to-end characterization of Llama 3 8B training under fully sharded data parallelism (FSDP) on an eight-GPU AMD Instinct\texttrademark{} MI300X node.
Our analysis reveals several previously underexplored bottlenecks and behaviors, such as memory determinism enabling higher, more stable GPU and memory frequencies. We identify several sources of inefficiencies, with frequency overhead (DVFS effects) being the single largest contributor to the gap between theoretical and observed performance, exceeding the impact of MFMA utilization loss, communication/computation overlap, and kernel launch overheads.
Overall, \name{} provides the first holistic, multi-granularity characterization of LLM training on AMD Instinct\texttrademark{} MI300X GPUs, yielding actionable insights for optimizing training frameworks, improving power-management strategies, and guiding future GPU architecture and system design.
\name{} code can be accessed at \href{https://anonymous.4open.science/r/chopper-7D49/README.md}{this anonymous GitHub repository}.
\end{abstract}

\section{Introduction}
\label{sec:Introduction}
AI has been the focus of the world for the last decade since AlexNet~\cite{krizhevsky2012imagenet}, and has been transforming many aspects of our life, such as healthcare~\cite{panch2019inconvenient}, social networking~\cite{hung2020social}, and entertainment~\cite{millington2019ai}.
Despite their transformative capability, AI workloads, especially generative AI, represented by LLMs, are extremely computationally expensive for both training and inference, due to large model size (e.g., up to 405 billion parameters for Llama 3~\cite{grattafiori2024llama}).
The wide adoption of the transformer in LLMs further stresses the computation, as its attention mechanism exhibits quadratic computational overheads with respect to the sequence length~\cite{vaswani2017attention}.

To accelerate AI workloads, a broad spectrum of hardware acceleration techniques have been proposed for inference, addressing distinct aspects (e.g., dataflow~\cite{eyeriss_paper, tpu_paper}, data format~\cite{carat_paper, fp8_paper1}, sparsity~\cite{bit_tactical_paper, bit_sparsity_paper}).
Regarding generative AI training, these inference-oriented hardware accelerators are usually not sufficient, due to three reasons.
First, generative AI is evolving rapidly in the size of both models and datasets (billions of model parameters and trillions of tokens in training datasets), requiring high scalability.
Inference accelerators are usually not designed to support thousands of interconnected devices in a distributed setting.
Second, generative AI supports a wide range of operations, requiring high flexibility.
However, most inference hardware is optimized for a limited number of operations.
Designed with scalability and flexibility, GPU systems are undoubtedly dominant in generative AI training~\cite{AMDInstinctWhereToBuy2025, gpu_power}.
In the last decade, to fuel the demand of AI, GPU systems have been deeply optimized from both software and hardware aspects.
Examples of software innovation include highly optimized linear algebra libraries~\cite{hipBLAS, cuBLAS},
kernel fusion~\cite{kernel_fusion_paper1, kernel_fusion_paper2}, and FlashAttention~\cite{flashattention_paper}, dynamic compilation~\cite{torch_compile_paper} for a single GPU,
as well as multi-GPU parallel computing with data parallelism~\cite{fsdp_paper},
pipeline parallelism~\cite{model_parallel_paper}, tensor parallelism~\cite{tensor_parallel_paper} and context parallelism~\cite{context_parallel_paper}.
Often, C3 (concurrent computation and communication) is leveraged to boost the performance of such parallelism strategies~\cite{tale_of_two_cs, c3}.
Examples of hardware optimizations include memory access management with a tensor memory accelerator~\cite{tensor_memory_accelerator_paper}, computation acceleration with tensor/matrix cores~\cite{tensor_core_paper, matrix_core_paper}, efficient data format with FP8~\cite{fp8_paper1, fp8_paper2}, and high bandwidth memory (HBM) integration~\cite{gpu_hbm_paper1, gpu_hbm_paper2}.
All these optimizations have offered one order of magnitude speed-up for GPU-based AI systems.

\minisection{Motivation.}
Despite significant end-to-end speed-ups in GPU systems, it remains unclear how close current systems are to their theoretical performance, and what prevents further performance gains.
This fact motivates this characterization work: \emph{how do these optimizations contribute to GPU performance in LLM training}?
Answering this provides multifaceted benefits for GPU-based LLM training in the long run.
Understanding the impact of optimizations on end-to-end performance can open up opportunities to not only better utilize existing systems, but also design future architectures.

\begin{table}[!t]
\centering
\caption{Comparison of characterization methodology.}
\begin{tabular}{lcccc}
    \toprule
    \multirow{2}{*}{\textbf{Work}} & \multicolumn{2}{c}{\textbf{Granularity}} & \multirow{2}{*}{\textbf{Insights}} & \multirow{2}{*}{\textbf{Tool}} \\
    \cmidrule{2-3}
    & \textbf{Application} & \textbf{Hardware} &  & \\
    \midrule
    Multi-level\cite{multi_level_analysis} & Workload & GPU microarch & Hardware  & No \\
    Bottleneck\cite{bottleneck_analysis_pred} & Kernel & GPU microarch & Hardware & No \\
    TC\cite{demystify_tensor_cores} & Kernel & Matrix core & Hardware & No \\
    TKLQT\cite{tklqt} & Operation & GPU \& CPU & Hardware & No  \\
    BERT\cite{demystify_bert} & Operation & GPU microarch & All & No  \\
    \name (ours) & All & All & All & Yes \\
    \bottomrule
\end{tabular}
\label{tab:prior_works_comparison}
\end{table}

\minisection{Proposal.}
In response to the above need to characterize LLM training on GPUs, we develop \name{}, a tool to automatically collect and analyze the kernel traces, as well as visualize the profiling results.
We highlight the comparison between our work and prior works in Table~\ref{tab:prior_works_comparison}.
First, this work offers the full characterization coverage across the application stack.
\name{} profiles at the GPU kernel level, but enables characterization at different granularities.
These granularities are the kernel, operation (which consists of one or more kernels), layer, phase (i.e., forward, backward, and optimizer), iteration, GPU, and the full workload.
Second, this work examines GPU resources in great detail at different hardware levels.
\name{} looks at a node of multiple GPUs, the microarchitecture of individual GPUs, and the CPU, facilitating a comprehensive analysis of the full system.
Third, this work derives insights from both software and hardware aspects, rather than most of the other works that only focus on a single aspect.
Fourth, this work open sources our developed \name{} tool, featuring public accessibility, while other works either offer no tools~\cite{multi_level_analysis, demystify_tensor_cores, demystify_bert} or do not release the tool~\cite{tklqt, bottleneck_analysis_pred}.
We evaluate a compute node of eight AMD Instinct\texttrademark{} MI300X GPUs for training Llama 3 8B under FSDP~\cite{fsdp_paper}.
Our contributions in this work are listed as below.
\begin{itemize}
    \item We characterize the performance of LLM training under FSDP in a multi-GPU system, and the characterization spans across varying granularity of both the application and the hardware.
    \item We develop a tool, \name{}, to automate the profiling and visualize the profiling results via architecture charts.
    \name{} is optimized for AMD Instinct\texttrademark{} MI300X GPUs, and can be easily extended to support other AMD GPUs, or even other GPU vendors.
    \item We draw insights by looking at throughput through the lens of the complex interplay between operation efficiency, operation overlap, power management decisions, launch overhead effects, and more.
    We also provide a breakdown of operation duration to quantify the gap between the actual and theoretical duration.
\end{itemize}

The remainder of this paper is organized as follows.
Section~\ref{sec:Background} reviews the background.
Then, Section~\ref{sec:Architecture} describes the \name framework.
Next, Section~\ref{sec:Implementation} and Section~\ref{sec:Evaluation} articulate the evaluation setup and results.
Finally, Section~\ref{sec:Conclusion} concludes this paper.

\section{Background}
\label{sec:Background}

\subsection{Llama}

\begin{figure}[!t]
	\centering
	\includegraphics[width=\linewidth]{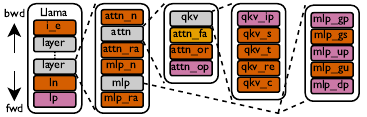}
	\caption{
		Diagram of operations in Llama.
		Colors of operation types match those used in Figure~\ref{fig:end_to_end}.
		i\_e is input embedding, ln is the final RMSNorm, lp is logits projection, attn/mlp\_n is attention/multi-layer perceptron (MLP) RMSNorm, attn/mlp\_ra is residual add, attn\_fa is FlashAttention, attn\_or is output reshape, attn\_op is output projection, qkv\_ip is the QKV (query, key, value) input projection, qkv\_s is split, qkv\_t is transpose, qkv\_re is rotary embedding, qkv\_c is contiguous memory copy, mlp\_gp is gate projection, mlp\_gs is silu, mlp\_up is up projection, mlp\_gu is gate-up elementwise multiply, and mlp\_dp is down projection.
	}
	\label{fig:transformer}
\end{figure}

Llama 3 are a set of publicly released LLMs~\cite{grattafiori2024llama} classified as foundation models, serving as the basis of an LLM platform.
Llama 3 is used for our pre-training benchmark, whose operations are detailed in Figure~\ref{fig:transformer}.

\subsection{Fully Sharded Data Parallelism}%
\label{sec:fsdp}

\begin{figure}[!t]
	\centering
	\includegraphics[width=\linewidth]{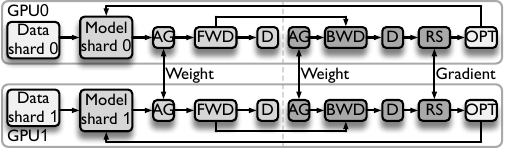}
	\caption{Overview of FSDP.
    Notation: all gather (AG); forward (FWD); backward (BWD); (D) delete extra model weights; reduce scatter (RS); optimize (OPT).
	}%
	\label{fig:fsdp_overview}
\end{figure}

While data parallelism has been widely used in AI training~\cite{data_parallel_paper}, generative AI models are prohibitively large to save on a single GPU, making data parallelism less efficient or even infeasible.
To address this limitation, FSDP is proposed~\cite{fsdp_paper}, as shown in Figure~\ref{fig:fsdp_overview}.
In addition to sharding the dataset, FSDP shards the model weights, gradients and optimizer states, with each shard being processed on one GPU,
enabling training larger AI models.
In Figure~\ref{fig:fsdp_overview}, FSDP starts with a shard of layer weights.
For the forward pass, each GPU collects shards from other GPUs using all gather to assemble the layer.
After the forward pass of the layer, the gathered shards are deleted.

The memory block corresponding to deleted layers can be re-used by the caching allocator for the upcoming layers to reduce memory use and fragmentation.
However, this is non-deterministic for FSDPv1, and the all gather may allocate a new block of memory before the layer is considered deleted, resulting in spikes of memory use~\cite{pytorch_blog}.
This behavior has been addressed with FSDPv2 by using per-parameter sharding.
However, this strategy introduces additional copies around communication collectives~\cite{pytorch_github_issue}, explored in Section~\ref{sec:kernel_launch}.

The computed activations from the forward pass will be used to compute gradients in the backward pass, before which another all gather is needed to collect all the weights from other GPUs.
After the backward pass, the weights are deleted again, and the gradients are summed and re-sharded across GPUs using reduce scatter.
Subsequently, the optimization step updates the weights locally on each GPU, after which the next iteration starts.
FSDP accomplishes this efficiently by utilizing C3.

\subsection{FlashAttention}
The attention mechanism is the core of transformer modules in LLMs~\cite{vaswani2017attention}.
Attention calculates the linear projection of query, key and value tokens,
followed by the product of query and key token matrices,
where the matrix size scales linearly with the sequence length.
Softmax is then applied to this product, which is further multiplied by the value matrix and a linear projection.
Each of these operations requires moving the data frequently between the off-chip GPU memory.
Such data movement leads to quadratic computational overheads with the sequence length, reaching more than $40\%$ of total runtime of transformers~\cite{softermax_paper}.

To address such quadratic overheads, FlashAttention is proposed to fuse the GPU kernels~\cite{flashattention_paper}.
FlashAttention splits the full input and computation into small tiles, and computes on one tile at a time, without moving intermediate results back to off-chip HBM, but at the cost of more memory accesses to on-chip SRAMs in GPUs.
FlashAttention also recomputes parts of the intermediate softmax values instead of storing all the results back to HBM, which is more expensive than computation.
To ensure numerical stability of softmax, FlashAttention uses online softmax with careful normalization to maintain stability while doing tile-by-tile operations~\cite{milakov2018online}.
This kernel fusion approach reduces the off-chip memory access significantly and achieves $7.6\times$ speed-up on attention~\cite{flashattention_paper}.

\subsection{AMD Matrix Core}
AMD matrix cores have been introduced since the first CDNA architecture~\cite{amd_cdna1_2020, amd_cdna2_2023, amd_cdna3_2023}, instantiated in AMD Instinct\texttrademark{} data center GPUs.
These matrix cores execute matrix fused multiply add (MFMA) instructions for general matrix multiply (GEMM) operations.
Over time, they have gained support
for a variety of data format for mixed precision computation,
from FP32
to FP16/BF16 and INT8/FP8.
These matrix cores are the fuel that powers
rocBLAS~\cite{amd_rocblas_latest}.
The AMD Instinct\texttrademark{} MI300X features 1,216 matrix cores, peaking at 1.3 BF16 peta floating-point operations per second (PFLOPS) at maximum frequency.
AMD provides tools to model the performance of these matrix cores~\cite{amd_matrix_instruction_calculator_2023}.

\section{\name Framework}
\label{sec:Architecture}

\subsection{Overview}
\begin{figure}[!t]
	\centering
	\includegraphics[width=\linewidth]{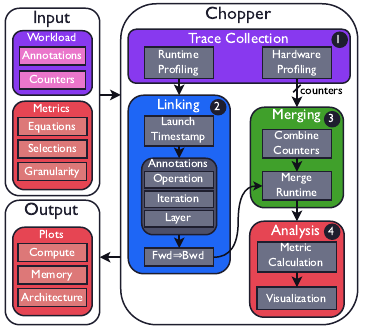}
	\caption{
		 Overview of the \name framework.
	}%
    \label{fig:framework_overview}
\end{figure}

Figure~\ref{fig:framework_overview} outlines our developed \name framework.
\name includes three modules for collecting, processing and analyzing the GPU execution traces.

\subsection{Trace Collection}

\subsubsection{Runtime Profiling}
The runtime profiling collects the execution traces, where the accurate timestamps of the executed GPU kernels and the launch process on the host CPU are recorded.
The collected trace also contains the mapping from forward kernels to backward kernels, since backward kernels are spawned from their forward counterparts, facilitating easy kernel recognition.
The collected trace further includes annotations for kernels, operations, layers, and iterations for the following trace alignment.

\subsubsection{Hardware Profiling}
The hardware profiling collects the target performance counters during GPU kernel execution.
Only a limited number of performance counters
can be collected at a time (e.g., we collect two or three at a time). However, collecting performance counters forces GPU kernels to be serialized.
This means performance counters cannot capture C3 overlap between GPU kernels, and cannot be used to extract valid timestamps as in runtime profiling.

\subsection{Trace Processing}

\subsubsection{Trace Alignment}
The trace alignment will combine and align the traces from runtime profiling and hardware profiling, such that the hardware counters
can be associated with high-level operations, layers and iterations.
Such alignment will facilitate the subsequent metric aggregation.

\subsection{Trace Analysis}

\subsubsection{Metric Aggregation}
Based on the aligned performance counters, we can aggregate the metrics by either directly reading individual performance counters, or derived metrics with equations based on multiple performance counters (e.g., calculating bandwidth from transferred bytes and kernel duration).
The aggregation can be constrained to a certain granularity (e.g., select specific GPUs, iterations, operation types, or individual operations).

\subsubsection{Visualization}
\name{} supports visualizing many different areas of the system from hardware utilization, to end-to-end performance.
The visualization capability goes beyond what is explored in Section~\ref{sec:Evaluation}.
Different visualizations use different aggregated metrics, and can be customized and filtered to the desired granularity.

\section{Experimental Setup}
\label{sec:Implementation}

\subsection{LLM Workload}
We use Llama 3 8B~\cite{grattafiori2024llama} as our workload.
We list the configuration in Table~\ref{tab:llm_config}.
Note that this model has group query attention enabled inherently~\cite{gqa_paper}.
In addition to the model configuration, we also sweep the batch size and sequence length.
We evaluate all configurations of batch size and sequence length that fit in the memory of the evaluated multi-GPU system.
This includes batch size (b) of one and sequence length (s) of 4k, denoted as b1s4.
Similar naming conventions are used for b2s4, b4s4, b1s8, and b2s8.

\begin{table}[!t]
\centering
\caption{Llama 3 8B model configuration.
\label{tab:llm_config}
}
\begin{tabular}{c|c|c|c}
    \toprule
    \textbf{Layer count} & \textbf{Token size} & \textbf{Hidden dim} & \textbf{Attn/KV heads} \\
    \midrule
    32 & 4,096  & 14,336 & 32/8 \\
    \bottomrule
\end{tabular}
\end{table}

\subsection{Training Framework}
In this work, we use a publicly available framework for LLM training, designed to benchmark pre-training for a single node with multiple GPUs on a synthesized dataset using PyTorch~\cite{amd_aig_aima_pytorch_benchmark_2025} with FSDP~\cite{fsdp_paper} and FSDPv2~\cite{fsdp2_tutorial}.
Note that neither additional parallelism nor advanced kernel fusion techniques (e.g., torch.compile~\cite{torch_compile_paper}) are adopted.
FlashAttention V2~\cite{dao2023flashattention} and the BF16 data format are used.

\subsection{Hardware System}
The compute node for training is composed of both host CPUs and accelerator GPUs~\cite{amd_hpcfund_hardware_2025}.
The host CPUs are two AMD EPYC\texttrademark{} 9684X CPUs, with 2.3 TB host memory in total.
There are a total of eight AMD Instinct\texttrademark{} MI300X GPUs, each with peak 1.3 PFLOPS and HBM of 192 GB capacity and 5.3 TB/s bandwidth~\cite{amd_cdna3_2023}.
Each pair of GPUs is connected via an AMD Infinity Fabric\texttrademark{} 128 GB/s bidirectional link, forming a fully connected eight-GPU system.
Each GPU is connected to the host CPU via a Gen 5 $\times$16 PCIe link.

\subsection{Profiling Tool}

We collect the performance counters using AMD rocprofv3~\cite{rocm_rocprofiler_sdk_2025}, a tool for advanced profiling and analytics for AMD hardware.
We collect the traces for LLM training using the PyTorch profiler which uses AMD roctracer~\cite{rocm_roctracer_2025}, a ROCm\texttrademark{} tracer callback/activity library for performance tracing AMD GPUs, under the hood.
20 training iterations are run, where the first 10 are warmup, and final 10 are sampled.
Training is run once with an optimizer phase at iteration 15 and once without.
Profiling metrics are derived using equations from rocprofiler-compute~\cite{rocprofiler-compute}.

\subsection{Setup Validation}
We validate the correctness of our training setup by comparing the reported token throughput and FLOPS for a similar model setup (Llama 3 8B with similar batch size and sequence length) and the same training setup (FSDP) on a similar system setup.
Our setup exhibits very close token throughput and FLOPS compared to prior works~\cite{semianalysis_mi300x_h100_h200_2024, amd_rocm_performance_results_2025}.

\section{Insights and Implications}%
\label{sec:Evaluation}

This section analyzes profiling results in a top-down manner, from end-to-end performance to operation runtime and variation, where we reason about the variation by looking at the relationship between runtime and C3 overlap. 
Afterward, the CPU behavior is analyzed and reasoned about in the context of launch overhead and core utilization.
Finally, we show the frequency and power, and use it to create a comprehensive breakdown, quantifying the gap between theoretical and actual performance. 
Throughout the evaluation, batch size may be referred to as $b$ and sequence length as $s$.

\subsection{End-to-End Performance Breakdown}

\begin{figure}[!t]
	\centering
	\includegraphics[width=\linewidth]{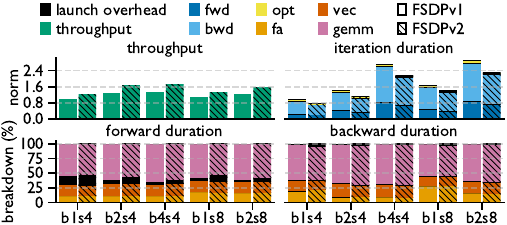}
	\caption{
		Median values across iterations and GPUs.
		Top row is normalized to b1s4 with FSDPv1.
		Duration is the sum of kernel duration.
		Launch overhead is the sum of bubbles between the previous and current kernel, and corresponds to the chunk or operation beneath it.
		Throughput is calculated with the maximum duration plus launch overhead across GPUs.
		Notation: batch size (b), sequence length in K tokens (s), forward pass (fwd), backward pass (bwd), optimization step (opt), FlashAttention (fa), vector operation (vec), matrix multiply (gemm).
	}%
	\label{fig:end_to_end}
\end{figure}

\subsubsection{Throughput Sensitivity to Batch Size/Sequence Length}

As shown in Figure~\ref{fig:end_to_end}, a batch size greater than one with sequence length 4K (b2s4, b4s4) 
achieves the highest throughput (token/sec), while batch size one (b1s4, b1s8) achieves the lowest throughput.
The significantly lower throughput indicates underutilization at batch size one.
We also observe slightly reduced throughput at a larger sequence length (b2s8).

\observation{
	Batch size one experiences severe underutilization (approximately 30\% lower throughput), regardless of the sequence length.
}

\subsubsection{Duration Breakdown---Phases \& Operation Types}%
\label{sec:end_to_end_dur}

The backward phase dominates training followed by the forward phase, with a marginal contribution from the optimizer phase.
Looking at the backward duration breakdown, FlashAttention occupies significantly more of the backward duration at batch size one than two (e.g., b1s4 versus b2s4).
It also occupies more of both forward and backward duration at a larger sequence length (e.g., b2s4 versus b2s8) as its duration scales with the square of sequence length.
This is the cause behind slightly lower throughput at a larger sequence length.
We also observe that GEMMs dominate training, occupying approximately 60\% of forward and backward duration.

\observation{
	Backward FlashAttention scales suboptimally, since it occupies a larger percentage of the backward breakdown at batch size one than two or four.
}

\subsubsection{Launch Overhead Across Batch Size/Sequence Length}

Launch overhead is the bubbles between kernels, illustrated in Figure~\ref{fig:launch_overhead_example}.
The optimizer phase and forward vector operations have the largest launch overheads in Figure~\ref{fig:end_to_end} (operations with high launch overhead will be explored in Section~\ref{sec:kernel_launch}).
Looking at iteration duration, the launch overhead is relatively constant across configurations, which causes it to occupy a larger percentage of duration for small batch sizes and sequence lengths (e.g., b1s4).

\observation{
	Launch overhead is more prominent in the forward and optimizer phases, and its percentage decreases with a larger batch size and sequence length.
}

\begin{figure}[!t]
    \centering
    \subfloat[
        GEMM and FA operation duration.
        Top row are forward operations, and bottom row are backward operations.
        Duration is normalized to the maximum of all configurations.%
    ]{\includegraphics[width=\linewidth]{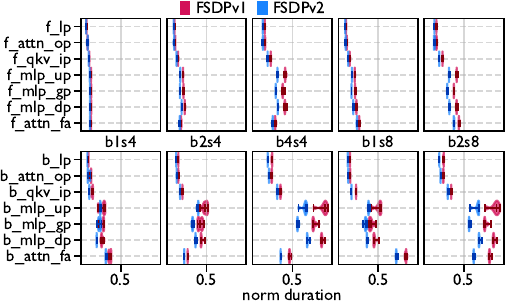}}%
    \label{fig:gemm_fa_time}
    \hfil
    \subfloat[
        Vec operation duration.
        Top row are forward operations, middle row are backward operations,
        and bottom row are optimizer operations.
        Duration is normalized to the maximum of all configurations.
    ]{\includegraphics[width=\linewidth]{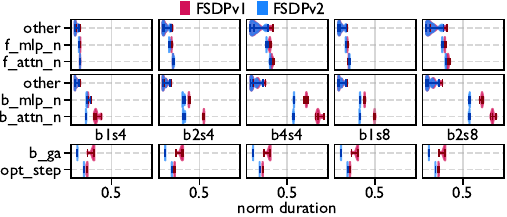}}%
    \label{fig:vec_time}
    \caption{
        Operation duration for different operator types and model configurations
        using FSDPv1 and v2. Duration is summed across layers and includes bubbles
        between the kernels of each operation.
    }%
    \label{fig:operation_time}
\end{figure}

\subsection{Operation Duration and Variation}%
\label{sec:operation_duration}

In this section, we identify the dominant operations in training, and variation in their duration.
Here, duration is defined as the sum of bubbles between, and runtime of all spawned kernels corresponding to a given operation.
The operation names are visualized and described in Figure~\ref{fig:transformer}, as well as f\_/b\_ to denote forward and backward, and b\_ga as the gradient accumulate operation going into the optimizer phase, and opt\_step as the optimizer step operation.

\subsubsection{GEMM}

All GEMMs scale with $b \cdot s$, with MLP dominating forward and backward, illustrated in Figure~\ref{fig:gemm_fa_time}.
These MLP GEMMs also exhibit significant variation in the backward phase.
In particular, the up projection (b\_mlp\_up) and gate projection (b\_mlp\_gp) have a distinct tail at a lower and higher duration, respectively.
Since the duration is aggregated across layers, data points are from iterations and GPUs.
This means a tail is likely from a few GPUs that are slower or faster than others (confirmed in Section~\ref{sec:comm_overlap}).

\subsubsection{FlashAttention}%
\label{sec:fa}

FlashAttention has comparable duration to the dominant MLP GEMMs in forward and backward, and begins to dominate at a larger sequence length (b1s8, b2s8) in Figure~\ref{fig:gemm_fa_time}.
While forward FlashAttention scales as expected with $b \cdot s^2$, backward FlashAttention has a lower duration at a batch size greater than one, despite performing more flops (i.e., b\_attn\_fa has a lower duration at b2s4 than b1s4, and b2s8 than b1s8).
This indicates that the backward FlashAttention implementation at batch size one is poorly optimized, as performing more flops should never decrease the duration.
This is why FlashAttention occupies a larger percentage of the backward breakdown of batch size one (b1s4, b1s8) in Figure~\ref{fig:end_to_end}, which also contributes to the underutilization observed.

\insight{
	Backward FlashAttention is poorly optimized for batch size one, as it has a lower duration at batch size two, despite performing more flops.
	This contributes to the underutilization at batch size one.
}

\rec{
	Leverage \name{} to visualize execution traces, identify, and fix implementation problems of Backward FlashAttention at batch size one.
}
\subsubsection{Vector}

RMSNorm operations (mlp\_n and attn\_n) dominate forward and backward in Figure~\ref{fig:vec_time}.
Both of these operations are identical (same computation, input, and output sizes) but have different durations, specifically for FSDPv1 in the backward phase. The major contributor of the increased duration for b\_attn\_n is communication overlap (explored in Section~\ref{sec:comm_overlap}).
We also observe the two operations for the optimizer phase (i.e., b\_ga and opt\_step) remain constant across sequence lengths and batch sizes, which makes sense as the shape of the model weights do not change across different batch sizes and sequence lengths.

\subsection{Communication \& Computation Overlap}%
\label{sec:comm_overlap}

In this section, we show the high correlation between communication overlap and computation duration, and how the overlap ratio and communication kernel duration varies across configurations.

\begin{figure}[!t]
	\centering
	\includegraphics[width=\linewidth]{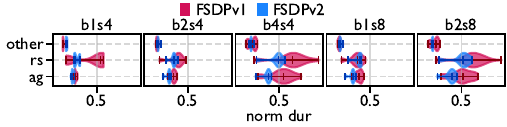}
	\caption{
		Per iteration duration of communication kernels.
		Duration is normalized to the maximum of all configurations.
	}%
	\label{fig:comm_kerns}
\end{figure}

\subsubsection{Communication Duration}

The main communication kernels are all gather (ag) and reduce scatter (rs).
Figure~\ref{fig:comm_kerns} shows the median communication duration scales with $b \cdot s$ like compute, represented by the iteration duration, while the tail remains relatively constant.
However, communication duration is a function of the hidden layers $H$ and number of GPU ranks $R$: $O(\frac{H}{R})$~\cite{tale_of_two_cs}.
Only the weights and gradients are communicated (which depend on $H$) and not activations (which depend on $H$, $b$,  and $s$).
Thus, communication duration should not change across $b$ and $s$.

\insight{
	While the tail communication duration follows theoretical trends (constant over $b$ and $s$), the median scales with compute time, indicating inefficiencies as the iteration duration grows.}

\rec{
	Modern systems need to better support the paradigm of C3 so it aligns closer to theoretical duration.
}

\begin{figure}[!t]
	\centering
	\includegraphics[width=\linewidth]{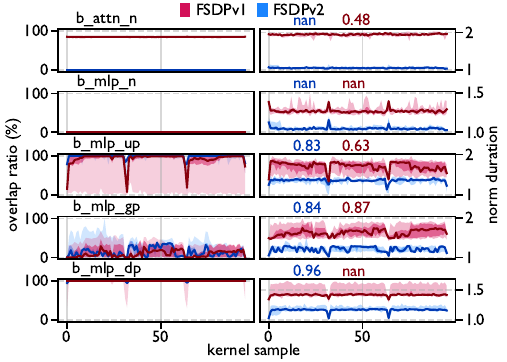}
	\caption{
		Overlap ratio versus duration of dominant vector and GEMM operations for three iterations at b2s4.
		The two numbers above each row on the right are the correlation between the overlap ratio and duration.
		The darker fill of the overlap ratio is the 25th to 75th quantile, and the light fill is the minimum to maximum.
	}%
	\label{fig:gemm_overlap}
\end{figure}

\subsubsection{GEMM Overlap}

Overlap has a high correlation with GEMM duration as shown by the correlation values above the backward MLP GEMMs in Figure~\ref{fig:gemm_overlap}.
Looking at b\_mlp\_up specifically, the fill shows that a few GPUs have an overlap ratio close to 0\%, while most have overlap close to 100\%.
This is reflected in the duration, with a few GPUs having approximately 15\% to 20\% lower duration than the median.
Other GEMMs show similar variation in overlap and duration, where one GPU has minimum overlap and little change in duration, while others have varying overlap and duration, as illustrated for f\_attn\_op in Figure~\ref{fig:per_gpu_overlap_variation}.
This confirms the tails in Figure~\ref{fig:gemm_fa_time} are from faster and slower GPUs, and not iterations.

\insight{
	Variation in overlap across GPUs contributes to variation in duration across GPUs.
}

\subsubsection{Vector Overlap}

While correlation is difficult to measure with constant overlap in the case of b\_attn\_n and b\_mlp\_n (low or nan values in Figure~\ref{fig:gemm_overlap}), we can compare the two operations since they perform identical computation.
By comparing the two operations, we observe b\_attn\_n is indeed impacted by its constant overlap of approximately 90\% for FSDPv1, since it has a larger duration than b\_mlp\_n which has 0\% overlap in Figure~\ref{fig:vec_time}.

\observation{
	Identical operations can have different durations as a result of their overlap ratio.
}

\subsubsection{FlashAttention Overlap}

\begin{figure}[!t]
	\centering
	\includegraphics[width=\linewidth]{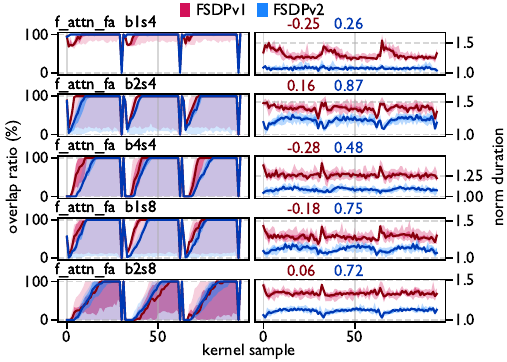}
	\caption{
		Overlap ratio versus duration for f\_attn\_fa under different model configurations.
		The fill is the same as Figure~\ref{fig:gemm_overlap}.
	}%
	\label{fig:overlap_sweep}
\end{figure}

For our configurations, only forward FlashAttention consistently experiences overlap.
Thus, the poor performance of backward FlashAttention observed in Section~\ref{sec:fa} cannot be attributed to overlap.
Figure~\ref{fig:overlap_sweep} shows how the overlap ratio of f\_attn\_fa changes as the batch size and sequence length increase, using the same fill as in Figure~\ref{fig:gemm_overlap}.
We observe that nearly all GPUs have approximately 100\% overlap at b1s4, but the fill and median value indicate that overlap decreases as the batch size and sequence length increase.
This makes sense, as FlashAttention duration scales with $b \cdot s^2$ while communication duration should remain constant.
Ultimately, smaller batch sizes experience more overlap, leading to more resource contention and underutilization.
This is another factor that helps to explain reduced throughput at small batch sizes in Figure~\ref{fig:end_to_end}.
We also observe lower correlation between overlap and duration for FlashAttention than we did for GEMMs.

\insight{
	Efficient overlap is especially important for smaller batch sizes and sequence lengths, which experience more overlap due to having shorter kernels. This causes more resource contention, affecting efficiency and throughput.
}

\begin{figure}[!t]
	\centering
	\includegraphics[width=\linewidth]{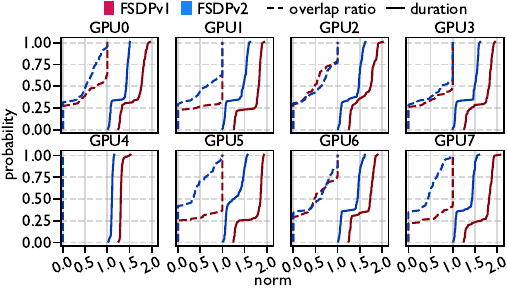}
	\caption{
		Cumulative distribution function (CDF) of overlap ratio versus duration of f\_attn\_op across eight GPUs for b2s4.
		Duration is normalized to the minimum value per GPU.
		Overlap ratio is normalized from zero to one.
	}%
	\label{fig:per_gpu_overlap_variation}
\end{figure}

\subsection{Kernel Launch Overheads---Causes and Implications}%
\label{sec:kernel_launch}

We previously observed significant launch overheads for the optimizer phase and forward vector operations in Figure~\ref{fig:end_to_end}. In this section, we will identify which specific operations were major contributors, and dissect launch overhead into preparation and call overheads.

\begin{figure}[!t]
	\centering
	\includegraphics[width=\linewidth]{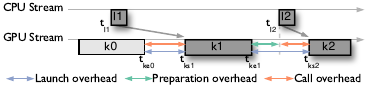}
	\caption{Example of launch overhead.
	}%
	\label{fig:launch_overhead_example}
\end{figure}

\begin{figure}[!t]
	\centering
	\includegraphics[width=\linewidth]{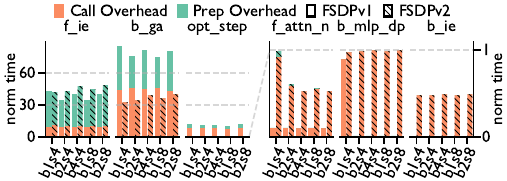}
	\caption{
		Mean Preparation and Call overhead for top operations, including bubbles between kernels within an operation.
	}%
	\label{fig:launch_overhead_scale}
\end{figure}

\begin{figure}[!t]
	\centering
	\includegraphics[width=\linewidth]{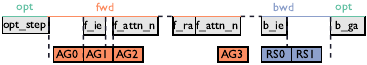}
	\caption{
		Example of filling and emptying the communication pipeline of all gathers and reduce scatters.
	}%
	\label{fig:fwd_bwd_launch_overhead_example}
\end{figure}

\subsubsection{Launch Overhead}

The launch overhead is visualized as in Figure~\ref{fig:launch_overhead_example} and formulated in Equation~\ref{eq:launch_overhead}, where $t_{\text{ks}_{i}}$ is the starting timestamp of a compute kernel $i$, $t_{\text{ke}_{i}}$ is the ending timestamp, and $t_{\text{l}_{i}}$ is kernel dispatch time.

\begin{equation}
	O_{\text{prep}} = \max(t_{\text{l}_{i}} - t_{\text{ke}_{i-1}}, 0)
	\label{eq:prep_overhead}
\end{equation}
\begin{equation}
	O_{\text{call}} = \min(t_{\text{ks}_{i}} - t_{\text{l}_{i}}, t_{\text{ks}_{i}} - t_{\text{ke}_{i-1}})
	\label{eq:call_overhead}
\end{equation}
\begin{equation}
	O_{\text{launch}} = O_{\text{prep}} + O_{\text{call}}
	\label{eq:launch_overhead}
\end{equation}

We consider the launch overhead as the bubbles between compute kernels, and ignore communication kernels.
This means that even if communication kernels are serialized and executed in the compute stream, they will be treated as bubbles and ignored, which can result in measuring a higher launch overhead (explored in Section~\ref{sec:call_overhead}).

\subsubsection{Preparation Overhead}

Preparation overhead is time that the CPU should have dispatched the next kernel but did not, marked green in Figure~\ref{fig:launch_overhead_example}.
Its value for a given kernel is zero if the CPU launch occurs before the end of the previous kernel, whereas a non-zero value indicates the kernel was launched ``too late.''
In theory, if a CPU is not the bottleneck, no preparation overhead is expected.
However, there are cases where the CPU does not need to dispatch a kernel so soon and is not the bottleneck.
We will also prove the CPU is not a bottleneck later by measuring core utilization in Section~\ref{sec:cpu_util}.

The two operations with large preparation overheads are f\_ie and opt\_step as shown in Figure~\ref{fig:launch_overhead_scale}.
Considering these operations happen at the start and end of an iteration respectively, we can reason that the preparation overhead is not indicative of a CPU bottleneck, and simply due to filling and emptying the pipeline of all gathers and reduce scatters as illustrated in Figure~\ref{fig:fwd_bwd_launch_overhead_example}.

\insight{
	Preparation overhead at the start and end of iterations arises from emptying and filling the pipeline with all gathers and reduce scatters, and does not indicate a CPU bottleneck.
}

\subsubsection{Call Overhead}%
\label{sec:call_overhead}

As expected, operations that occur while filling and emptying the communication pipeline, f\_ie and b\_ga, have the highest call overheads.
Additionally, the opt\_step operation has high call overhead, which occurs during the optimizer phase.
This operation has many small vector kernels with large bubbles between them.
However, these bubbles are significantly reduced going from FSDPv1 to FSDPv2.

Other operations have much lower call overhead compared to the three previously mentioned ones, illustrated by the gray dotted line connecting the left subplot's y-scale to the right in Figure~\ref{fig:launch_overhead_scale}.
For FSDPv1, f\_attn\_n is the only operation with notable call overhead. This is likely a result of the operation occurring while the initial all gathers are running consecutively, causing resource contention which prevents f\_attn\_n from executing earlier as illustrated in Figure~\ref{fig:fwd_bwd_launch_overhead_example}.
Interestingly, we observe more call overhead in FSDPv2 when comparing it to FSDPv1.
This is because FSDPv2 serializes copy kernels with the compute stream (as a result of per-parameter sharding explained in Section~\ref{sec:fsdp}) before f\_attn\_n, b\_mlp\_dp, and b\_ie, which appears as launch overhead.
Other operations have negligible call overhead and are omitted from Figure~\ref{fig:launch_overhead_scale}.

\observation{
	FSDPv2 serializes more copy operations, yet achieves significantly higher throughput than FSDPv1.
}

\subsubsection{End-to-end}

Now that operations with high launch overhead are identified and the reasons behind it are clear, we can explain the impact on the end-to-end performance in Figure~\ref{fig:end_to_end}.
The reason launch overhead occupies such a large portion of the forward vector duration is because f\_ie and f\_attn\_n are vector operations which occur while the pipeline is being filled with initial all gathers.
Since communication duration does not scale with batch size and sequence length, we observe its impact/percentage decreases as the batch size and sequence length become larger.
This is a third reason why we observe underutilization at small batch sizes.

\insight{
	The impact of launch overhead diminishes as the batch size and sequence length increase.
}
\rec{
	Typical graph launch optimizations focus on intra-iteration launch overheads~\cite{intra_iter_launch}, while inter-iteration overhead dominates. These techniques should be augmented to consider such overheads.
}

\begin{figure}[!t]
	\centering
	\includegraphics[width=\linewidth]{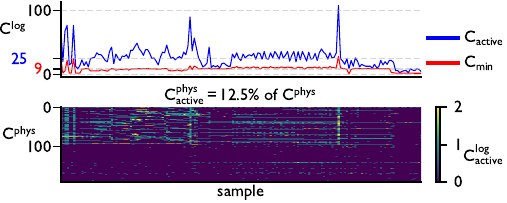}
	\caption{
		Minimum and Active cores in the top row, with logical to physical core mapping in the bottom row.
		FSDPv2 with b2s4 and no optimizer phase.
	}%
	\label{fig:cpu_phys_log}
\end{figure}

\subsection{CPU Utilization}%
\label{sec:cpu_util}

Based on the low preparation overhead (not a bottleneck) observed in Section~\ref{sec:kernel_launch} (aside from filling and emptying the pipeline), we expect the CPU to be underutilized.
We can confirm this by profiling the CPU during training and measuring the core utilization as illustrated in Figure~\ref{fig:cpu_phys_log}.

\subsubsection{Logical Cores}

We measure the logical core utilization in the top row of Figure~\ref{fig:cpu_phys_log} using $C_{\text{active}}$ which is the number of active cores with non-zero utilization, and $C_{\text{min}}$ which is the theoretical lower bound on active cores, and $N$ is the number of logical cores.

\begin{equation}
C_{\text{active}} = \sum_{i=1}^{N} [\text{Util}_i > 0],\quad
[{P}] =
\begin{cases}
1 & \text{if } P \text{ is true}; \\
0 & \text{otherwise}.
\end{cases}
\end{equation}

\begin{equation}
C_{\text{min}} = \sum_{i=1}^{N} \frac{\text{Util}_i}{100}
\end{equation}

There is a median of 25 active cores, despite a lower bound of nine minimum cores.
This suggests that the number of active cores could be reduced by more than two times with higher core utilization,
which would increase the number of idle cores.
Doing so creates an opportunity for power-gating, or power-sloshing to reallocate power to the GPUs.

\subsubsection{Physical Cores}

Our hardware platform has simultaneous multithreading (SMT) enabled.
This means that two logical cores can be mapped to the same physical core.
However, we can see this rarely happens in Figure~\ref{fig:cpu_phys_log}, with the heatmap rarely having yellow data points.
Only $12.5\%$ of physical cores have one or more active logical cores mapped over the course of training.
This indicates that the CPU is heavily underutilized, even with the active core count more than double the lower bound.

\insight{
	The CPU is heavily underutilized in LLM training, despite the fact that active cores are more than double the lower bound.
}

\rec{
	Future LLM training systems do not need as many active CPU cores.
    System designers can slosh the CPU power to GPUs without impacting training performance.
}

\begin{figure}[!t]
	\centering
	\includegraphics[width=\linewidth]{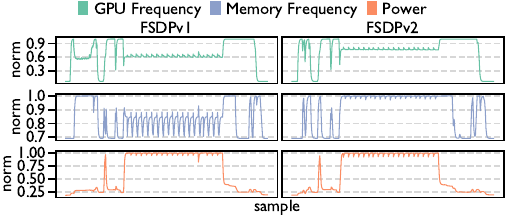}
	\caption{Average frequency and power for FSDPv1 and FSDPv2 with b2s4 and no optimizer phase.}%
	\label{fig:average_freq_power}
\end{figure}

\begin{figure*}[t!]
	\centering
	\includegraphics[width=\linewidth]{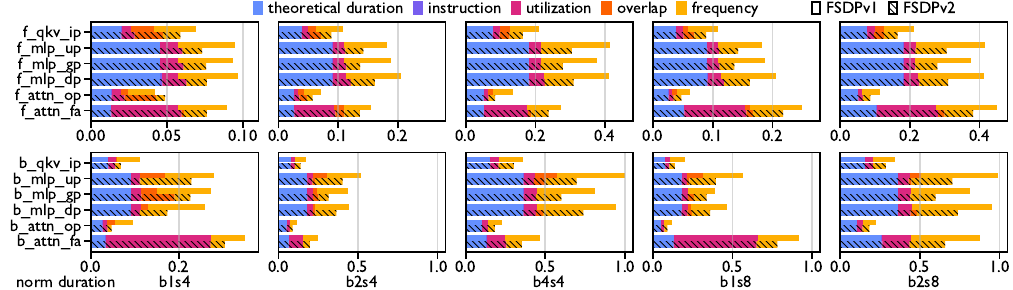}
	\caption{Overhead breakdown for GEMMs and FlashAttention.}%
	\label{fig:gemm_fa_overhead}
\end{figure*}

\subsection{Frequency and Power}

In Section~\ref{sec:comm_overlap}, we observed that operations tended to have lower runtime for FSDPv2, even if both FSDPv1 and FSDPv2 had 0\% overlap (e.g., b\_mlp\_n in Figure~\ref{fig:gemm_overlap} and f\_attn\_op for GPU4 in Figure~\ref{fig:per_gpu_overlap_variation}).
However, the model does not change, and the primary compute kernels are the same.
Thus, frequency is the most likely factor to explain a uniformly lower duration under FSDPv2 for all operations as illustrated in Figure~\ref{fig:operation_time}.
Indeed, FSDPv2 achieves approximately $20\%$ higher GPU and memory clock frequency with significantly less variation, and a nearly identical power signature as FSDPv1 in Figure~\ref{fig:average_freq_power}.
Because FSDPv2 introduces more deterministic memory allocation behavior (Section~\ref{sec:fsdp}), we hypothesize that this reduces HBM power variability, increasing efficiency and allowing the GPU and memory to sustain higher clock frequencies under the same power constraints.

\observation{
	FSDPv2 is more power efficient, allowing the average GPU frequency to be approximately 25\% higher with no change in power.
}

\subsection{Duration Breakdown---the Gap Between Theoretical and Actual Performance}

Now that all we have a broad view of phenomenon affecting performance, we can make an overhead breakdown to show exactly which factors are limiting the theoretical performance, and the contribution each factor has as illustrated in Figure~\ref{fig:gemm_fa_overhead}.

\subsubsection{Theoretical duration}

This is obtained by dividing an operation's theoretical flops ($F_{\text{gemm}}$) by the peak FLOPS ($\text{TPT}_{\text{peak}}$).
\begin{equation}
	\label{eq:thr_duration}
	D_{\text{thr}} = \frac{F_{\text{gemm}}}{\text{TPT}_{\text{peak}}}
\end{equation}

\subsubsection{Instruction overhead}

This is the overhead of performing more flops than needed (padding), and calculated as the ratio of an operation's performed flops ($F_{\text{perf}}$) to theoretical flops ($F_{\text{gemm}}$).
\begin{equation}
	\label{eq:inst_overhead}
	\text{\text{Ovr}}_{\text{inst}} = \frac{F_{\text{perf}}}{F_{\text{gemm}}}
\end{equation}

Instruction overhead is rare, only visible for f\_mlp\_dp at b1s4.

\subsubsection{Utilization overhead}

This is the overhead of MFMA cores not running at 100\% utilization, and calculated as the inverse of MFMA utilization ($\text{MFMA}_{\text{util}}$).
\begin{equation}
	\label{eq:util_overhead}
	\text{\text{Ovr}}_{\text{util}} = \frac{1}{\text{MFMA}_{\text{util}}}
\end{equation}

Utilization overhead appears particularly high for FlashAttention, because it has to perform vector operations as well as GEMM.
It is very similar between FSDPv1 and FSDPv2, confirming the same compute kernels are being ran.

\subsubsection{Overlap overhead}

This is an approximation of the overhead from C3 resource contention.
It is extracted from a CDF like Figure~\ref{fig:per_gpu_overlap_variation} and calculated as the ratio of an operation's duration at 50\% of the overlap ($D_{\text{50\%}}$) to its duration at 0\% ($D_{\text{0\%}}$).
\begin{equation}
	\label{eq:overlap_overhead}
	\text{\text{Ovr}}_{\text{overlap}} = \frac{D_{\text{50\%}}}{D_{\text{0\%}}}
\end{equation}

As expected, overlap overhead decreases as the batch size and sequence length grow. FSDPv2 has similar overlap overhead to FSDPv1, sometimes introducing more overlap overhead, but most of the time decreasing it which we expect based on its lower median communication kernel duration in Figure~\ref{fig:comm_kerns}.

\subsubsection{Frequency overhead}

This is the overhead from running below peak frequency and calculated with a few steps.
First, peak clock duration ($D_{\text{peak}}$) is calculated by dividing the GPU cycles an operation took ($C_{\text{gpu}}$) by the GPU's peak clock frequency ($\text{Freq}_{\text{peak}}$).
Next, the ratio of actual duration ($D_{\text{act}}$) to peak clock duration is the temporary frequency overhead.
Finally, Overlap overhead is divided to adjust the frequency overhead to more accurately represent overhead from dynamic voltage and frequency scaling (DVFS).
\begin{equation}
	\label{eq:freq_overhead}
	D_{\text{peak}} = \frac{C_{\text{gpu}}}{\text{Freq}_{\text{peak}}}, \quad \text{\text{Ovr}}_{\text{freq}} = \frac{D_{\text{act}}}{D_{\text{peak}}}/\text{\text{Ovr}}_{\text{overlap}}
\end{equation}

This overhead dominates, in particular for GEMM.
It is also the biggest difference between FSDPv1 and FSDPv2, confirmed by Figure~\ref{fig:average_freq_power}.

\insight{
	Frequency overhead dominates, and is the biggest improvement from FSDPv1 to FSDPv2.
}
\rec{
	Frequency should be a principal component of profiling, especially when comparing training frameworks.
	Furthermore, power management firmware tuning and optimization is necessary as the same workload manifests different frequency decisions across frameworks.
}

\section{Conclusion}
\label{sec:Conclusion}
Modern GPU systems are heavily shaped by AI, especially generative AI.
Given the complexity of modern GPU hardware and software stacks, we aim to understand how individual components collectively shape end-to-end LLM training performance.
In this work, we characterize the performance of AMD Instinct\texttrademark{} MI300X GPUs when training Llama 3 8B model under FSDP in a single-node, eight-GPU system.
We develop \name{} to automate the trace collection and analysis, as well as the result visualization.
We derive insights based on operation variation and overlap, CPU behavior and more, shedding light on the optimization of both current and future GPU architecture and systems.

\section{Acknowledgment}
This work was sponsored by the Funding for Academic Research program (gift funding) under the AMD University Program.
Access to GPUs was provided by the AMD University Program AI \& HPC Cluster and the AMD Developer Cloud.

AMD, AMD Instinct, AMD EPYC, AMD Infinity Fabric, and combinations thereof are trademarks of Advanced Micro Devices, Inc.
Other product names used in this publication are for identification purposes only and may be trademarks of their respective companies.

\bibliographystyle{IEEEtran}
\bibliography{reference}

\begin{thebibliography}{10}
\providecommand{\url}[1]{#1}
\csname url@samestyle\endcsname
\providecommand{\newblock}{\relax}
\providecommand{\bibinfo}[2]{#2}
\providecommand{\BIBentrySTDinterwordspacing}{\spaceskip=0pt\relax}
\providecommand{\BIBentryALTinterwordstretchfactor}{4}
\providecommand{\BIBentryALTinterwordspacing}{\spaceskip=\fontdimen2\font plus
\BIBentryALTinterwordstretchfactor\fontdimen3\font minus
  \fontdimen4\font\relax}
\providecommand{\BIBforeignlanguage}[2]{{%
\expandafter\ifx\csname l@#1\endcsname\relax
\typeout{** WARNING: IEEEtran.bst: No hyphenation pattern has been}%
\typeout{** loaded for the language `#1'. Using the pattern for}%
\typeout{** the default language instead.}%
\else
\language=\csname l@#1\endcsname
\fi
#2}}
\providecommand{\BIBdecl}{\relax}
\BIBdecl

\bibitem{krizhevsky2012imagenet}
A.~Krizhevsky, I.~Sutskever, and G.~E. Hinton, ``{Imagenet Classification with
  Deep Convolutional Neural Networks},'' \emph{Advances in neural information
  processing systems}, vol.~25, 2012.

\bibitem{panch2019inconvenient}
T.~Panch, H.~Mattie, and L.~A. Celi, ``{The “Inconvenient Truth” About AI
  in Healthcare},'' \emph{NPJ digital medicine}, vol.~2, no.~1, pp. 1--3, 2019.

\bibitem{hung2020social}
M.~Hung, E.~Lauren, E.~S. Hon, W.~C. Birmingham, J.~Xu, S.~Su, S.~D. Hon,
  J.~Park, P.~Dang, and M.~S. Lipsky, ``{Social Network Analysis of COVID-19
  Sentiments: Application of Artificial Intelligence},'' \emph{Journal of
  medical Internet research}, vol.~22, no.~8, p. e22590, 2020.

\bibitem{millington2019ai}
I.~Millington, \emph{{AI for Games}}.\hskip 1em plus 0.5em minus 0.4em\relax
  CRC Press, 2019.

\bibitem{grattafiori2024llama}
A.~.~M. Llama~Team, ``{The Llama 3 Herd of Models},'' \emph{arXiv preprint
  arXiv:2407.21783}, 2024.

\bibitem{vaswani2017attention}
A.~Vaswani, N.~Shazeer, N.~Parmar, J.~Uszkoreit, L.~Jones, A.~N. Gomez,
  {\L}.~Kaiser, and I.~Polosukhin, ``{Attention Is All You Need},''
  \emph{Advances in neural information processing systems}, vol.~30, 2017.

\bibitem{eyeriss_paper}
Y.-H. Chen, J.~Emer, and V.~Sze, ``{Eyeriss: A Spatial Architecture for
  Energy-Efficient Dataflow for Convolutional Neural Networks},'' in
  \emph{International Symposium on Computer Architecture}, 2016.

\bibitem{tpu_paper}
N.~P. Jouppi, C.~Young, N.~Patil, D.~Patterson, G.~Agrawal, R.~Bajwa, S.~Bates,
  S.~Bhatia, N.~Boden, A.~Borchers, R.~Boyle, P.-l. Cantin, C.~Chao, C.~Clark,
  J.~Coriell, M.~Daley, M.~Dau, J.~Dean, B.~Gelb, T.~V. Ghaemmaghami,
  R.~Gottipati, W.~Gulland, R.~Hagmann, C.~R. Ho, D.~Hogberg, J.~Hu, R.~Hundt,
  D.~Hurt, J.~Ibarz, A.~Jaffey, A.~Jaworski, A.~Kaplan, H.~Khaitan,
  D.~Killebrew, A.~Koch, N.~Kumar, S.~Lacy, J.~Laudon, J.~Law, D.~Le, C.~Leary,
  Z.~Liu, K.~Lucke, A.~Lundin, G.~MacKean, A.~Maggiore, M.~Mahony, K.~Miller,
  R.~Nagarajan, R.~Narayanaswami, R.~Ni, K.~Nix, T.~Norrie, M.~Omernick,
  N.~Penukonda, A.~Phelps, J.~Ross, M.~Ross, A.~Salek, E.~Samadiani, C.~Severn,
  G.~Sizikov, M.~Snelham, J.~Souter, D.~Steinberg, A.~Swing, M.~Tan,
  G.~Thorson, B.~Tian, H.~Toma, E.~Tuttle, V.~Vasudevan, R.~Walter, W.~Wang,
  E.~Wilcox, and D.~H. Yoon, ``{In-Datacenter Performance Analysis of A Tensor
  Processing Unit},'' in \emph{International Symposium on Computer
  Architecture}, 2017.

\bibitem{carat_paper}
Z.~Pan, J.~San~Miguel, and D.~Wu, ``{Carat: Unlocking Value-Level Parallelism
  for Multiplier-Free GEMMs },'' in \emph{International Conference on
  Architectural Support for Programming Languages and Operating Systems}, 2024.

\bibitem{fp8_paper1}
P.~Micikevicius, D.~Stosic, N.~Burgess, M.~Cornea, P.~Dubey, R.~Grisenthwaite,
  S.~Ha, A.~Heinecke, P.~Judd, J.~Kamalu \emph{et~al.}, ``{Fp8 Formats for Deep
  Learning},'' \emph{arXiv preprint arXiv:2209.05433}, 2022.

\bibitem{bit_tactical_paper}
A.~Delmas~Lascorz, P.~Judd, D.~M. Stuart, Z.~Poulos, M.~Mahmoud, S.~Sharify,
  M.~Nikolic, K.~Siu, and A.~Moshovos, ``{Bit-Tactical: A Software/Hardware
  Approach to Exploiting Value and Bit Sparsity in Neural Networks},'' in
  \emph{International Conference on Architectural Support for Programming
  Languages and Operating Systems}, 2019.

\bibitem{bit_sparsity_paper}
H.~Lu, L.~Chang, C.~Li, Z.~Zhu, S.~Lu, Y.~Liu, and M.~Zhang, ``{Distilling
  Bit-Level Sparsity Parallelism for General Purpose Deep Learning
  Acceleration},'' in \emph{International Symposium on Microarchitecture},
  2021.

\bibitem{AMDInstinctWhereToBuy2025}
{Advanced Micro Devices, Inc.}, ``{{AMD Instinct™ Solutions: Where to Buy
  Accelerators}},'' \url{
  https://www.amd.com/en/where-to-buy/accelerators/instinct.html }, 2025,
  accessed on 2025-06-07.

\bibitem{gpu_power}
\BIBentryALTinterwordspacing
Continuum, ``{NVIDIA GB200 NVL72},'' 2025, accessed on April 13, 2025.
  [Online]. Available:
  \url{https://training.continuumlabs.ai/infrastructure/servers-and-chips/nvidia-gb200-nvl72}
\BIBentrySTDinterwordspacing

\bibitem{hipBLAS}
{Advanced Micro Devices, Inc.}, ``{hipBLAS: ROCm BLAS marshalling library},''
  \url{https://github.com/ROCm/hipBLAS}, 2024, accessed: 2025-06-08.

\bibitem{cuBLAS}
{NVIDIA Corporation}, ``{cuBLAS: GPU-accelerated Basic Linear Algebra
  Subprograms},'' \url{https://developer.nvidia.com/cublas}, 2024, accessed:
  2025-06-08.

\bibitem{kernel_fusion_paper1}
A.~Li, B.~Zheng, G.~Pekhimenko, and F.~Long, ``{Automatic Horizontal Fusion for
  GPU Kernels},'' in \emph{2022 IEEE/ACM International Symposium on Code
  Generation and Optimization (CGO)}, 2022, pp. 14--27.

\bibitem{kernel_fusion_paper2}
W.~Niu, J.~Guan, Y.~Wang, G.~Agrawal, and B.~Ren, ``{Dnnfusion: Accelerating
  Deep Neural Networks Execution with Advanced Operator Fusion},'' in
  \emph{Proceedings of the 42nd ACM SIGPLAN International Conference on
  Programming Language Design and Implementation}, 2021, pp. 883--898.

\bibitem{flashattention_paper}
T.~Dao, D.~Y. Fu, S.~Ermon, A.~Rudra, and C.~R\'{e}, ``{FLASHATTENTION: Fast
  and Memory-Efficient Exact Attention with IO-Awareness},'' in
  \emph{Proceedings of the 36th International Conference on Neural Information
  Processing Systems}, ser. NIPS '22.\hskip 1em plus 0.5em minus 0.4em\relax
  Red Hook, NY, USA: Curran Associates Inc., 2022.

\bibitem{torch_compile_paper}
\BIBentryALTinterwordspacing
J.~Ansel, E.~Yang, H.~He, N.~Gimelshein, A.~Jain, M.~Voznesensky, B.~Bao,
  P.~Bell, D.~Berard, E.~Burovski, G.~Chauhan, A.~Chourdia, W.~Constable,
  A.~Desmaison, Z.~DeVito, E.~Ellison, W.~Feng, J.~Gong, M.~Gschwind, B.~Hirsh,
  S.~Huang, K.~Kalambarkar, L.~Kirsch, M.~Lazos, M.~Lezcano, Y.~Liang,
  J.~Liang, Y.~Lu, C.~K. Luk, B.~Maher, Y.~Pan, C.~Puhrsch, M.~Reso,
  M.~Saroufim, M.~Y. Siraichi, H.~Suk, S.~Zhang, M.~Suo, P.~Tillet, X.~Zhao,
  E.~Wang, K.~Zhou, R.~Zou, X.~Wang, A.~Mathews, W.~Wen, G.~Chanan, P.~Wu, and
  S.~Chintala, ``{PyTorch 2: Faster Machine Learning Through Dynamic Python
  Bytecode Transformation and Graph Compilation},'' in \emph{Proceedings of the
  29th ACM International Conference on Architectural Support for Programming
  Languages and Operating Systems, Volume 2}, ser. ASPLOS '24.\hskip 1em plus
  0.5em minus 0.4em\relax New York, NY, USA: Association for Computing
  Machinery, 2024, p. 929–947. [Online]. Available:
  \url{https://doi.org/10.1145/3620665.3640366}
\BIBentrySTDinterwordspacing

\bibitem{fsdp_paper}
\BIBentryALTinterwordspacing
Y.~Zhao, A.~Gu, R.~Varma, L.~Luo, C.-C. Huang, M.~Xu, L.~Wright,
  H.~Shojanazeri, M.~Ott, S.~Shleifer, A.~Desmaison, C.~Balioglu, P.~Damania,
  B.~Nguyen, G.~Chauhan, Y.~Hao, A.~Mathews, and S.~Li, ``{PyTorch FSDP:
  Experiences on Scaling Fully Sharded Data Parallel},'' \emph{Proc. VLDB
  Endow.}, vol.~16, no.~12, p. 3848–3860, Aug. 2023. [Online]. Available:
  \url{https://doi.org/10.14778/3611540.3611569}
\BIBentrySTDinterwordspacing

\bibitem{model_parallel_paper}
Z.~Jia, M.~Zaharia, and A.~Aiken, ``{Beyond Data and Model Parallelism for Deep
  Neural Networks.}'' \emph{Proceedings of Machine Learning and Systems},
  vol.~1, pp. 1--13, 2019.

\bibitem{tensor_parallel_paper}
M.~Shoeybi, M.~Patwary, R.~Puri, P.~LeGresley, J.~Casper, and B.~Catanzaro,
  ``{Megatron-LM: Training Multi-Billion Parameter Language Models Using Model
  Parallelism},'' \emph{arXiv preprint arXiv:1909.08053}, 2019.

\bibitem{context_parallel_paper}
A.~Yang, J.~Yang, A.~Ibrahim, X.~Xie, B.~Tang, G.~Sizov, J.~Reizenstein,
  J.~Park, and J.~Huang, ``{Context Parallelism for Scalable Million-Token
  Inference},'' \emph{arXiv preprint arXiv:2411.01783}, 2024.

\bibitem{tale_of_two_cs}
S.~Pati, S.~Aga, M.~Islam, N.~Jayasena, and M.~D. Sinclair, ``{Tale of Two Cs:
  Computation vs. Communication Scaling for Future Transformers on Future
  Hardware},'' in \emph{IEEE International Symposium on Workload
  Characterization}, 2023.

\bibitem{c3}
A.~Agrawal, S.~Aga, S.~Pati, and M.~Islam, ``{ConCCL: Optimizing ML Concurrent
  Computation and Communication with GPU DMA Engines},'' in \emph{2025 IEEE
  International Symposium on Performance Analysis of Systems and Software
  (ISPASS)}, 2025.

\bibitem{tensor_memory_accelerator_paper}
\BIBentryALTinterwordspacing
N.~Meseguer, Y.~Sun, M.~Pellauer, J.~L. Abell\'{a}n, and M.~E. Acacio, ``{ACTA:
  Automatic Configuration of the Tensor Memory Accelerator for High-End
  GPUs},'' in \emph{Proceedings of the 17th Workshop on General Purpose
  Processing Using GPU}, ser. GPGPU '25.\hskip 1em plus 0.5em minus 0.4em\relax
  New York, NY, USA: Association for Computing Machinery, 2025, p. 21–27.
  [Online]. Available: \url{https://doi.org/10.1145/3725798.3725802}
\BIBentrySTDinterwordspacing

\bibitem{tensor_core_paper}
S.~Markidis, S.~W. Der~Chien, E.~Laure, I.~B. Peng, and J.~S. Vetter, ``{Nvidia
  Tensor Core Programmability, Performance \& Precision},'' in \emph{2018 IEEE
  international parallel and distributed processing symposium workshops
  (IPDPSW)}.\hskip 1em plus 0.5em minus 0.4em\relax IEEE, 2018, pp. 522--531.

\bibitem{matrix_core_paper}
G.~Schieffer, D.~A. De~Medeiros, J.~Faj, A.~Marathe, and I.~Peng, ``{On the
  Rise of Amd Matrix Cores: Performance, Power Efficiency, And
  Programmability},'' in \emph{2024 IEEE International Symposium on Performance
  Analysis of Systems and Software (ISPASS)}.\hskip 1em plus 0.5em minus
  0.4em\relax IEEE, 2024, pp. 132--143.

\bibitem{fp8_paper2}
A.~Kuzmin, M.~Van~Baalen, Y.~Ren, M.~Nagel, J.~Peters, and T.~Blankevoort,
  ``{Fp8 Quantization: The Power of the Exponent},'' \emph{Advances in Neural
  Information Processing Systems}, vol.~35, pp. 14\,651--14\,662, 2022.

\bibitem{gpu_hbm_paper1}
J.~Choquette, ``{Nvidia Hopper Gpu: Scaling Performance},'' in \emph{2022 IEEE
  Hot Chips 34 Symposium (HCS)}.\hskip 1em plus 0.5em minus 0.4em\relax IEEE
  Computer Society, 2022, pp. 1--46.

\bibitem{gpu_hbm_paper2}
A.~Smith, E.~Chapman, C.~Patel, R.~Swaminathan, J.~Wuu, T.~Huang, W.~Jung,
  A.~Kaganov, H.~McIntyre, and R.~Mangaser, ``{11.1 AMD InstinctTM MI300 Series
  Modular Chiplet Package--HPC and AI Accelerator for Exa-Class Systems},'' in
  \emph{2024 IEEE International Solid-State Circuits Conference (ISSCC)},
  vol.~67.\hskip 1em plus 0.5em minus 0.4em\relax IEEE, 2024, pp. 490--492.

\bibitem{multi_level_analysis}
P.~Delestrac, D.~Battacharjee, S.~Yang, D.~Moolchandani, F.~Catthoor,
  L.~Torres, and D.~Novo, ``{Multi-Level Analysis of GPU Utilization in ML
  Training Workloads},'' in \emph{2024 Design, Automation \& Test in Europe
  Conference \& Exhibition (DATE)}, 2024, pp. 1--6.

\bibitem{bottleneck_analysis_pred}
S.~Madougou, A.~L. Varbanescu, C.~De~Laat, and R.~Van~Nieuwpoort, ``{A Tool for
  Bottleneck Analysis and Performance Prediction for GPU-Accelerated
  Applications},'' in \emph{2016 IEEE International Parallel and Distributed
  Processing Symposium Workshops (IPDPSW)}, 2016, pp. 641--652.

\bibitem{demystify_tensor_cores}
D.~Yan, W.~Wang, and X.~Chu, ``{Demystifying Tensor Cores to Optimize
  Half-Precision Matrix Multiply},'' in \emph{2020 IEEE International Parallel
  and Distributed Processing Symposium (IPDPS)}, 2020, pp. 634--643.

\bibitem{tklqt}
P.~Vellaisamy, T.~Labonte, S.~Chakraborty, M.~Turner, S.~Sury, and J.~P. Shen,
  ``{Characterizing and Optimizing LLM Inference Workloads on CPU-GPU Coupled
  Architectures},'' \emph{arXiv preprint arXiv:2504.11750}, 2025.

\bibitem{demystify_bert}
S.~Pati, S.~Aga, N.~Jayasena, and M.~D. Sinclair, ``{Demystifying BERT: System
  Design Implications},'' in \emph{2022 IEEE International Symposium on
  Workload Characterization (IISWC)}, 2022, pp. 296--309.

\bibitem{data_parallel_paper}
\BIBentryALTinterwordspacing
S.~Li, Y.~Zhao, R.~Varma, O.~Salpekar, P.~Noordhuis, T.~Li, A.~Paszke,
  J.~Smith, B.~Vaughan, P.~Damania, and S.~Chintala, ``{PyTorch Distributed:
  Experiences on Accelerating Data Parallel Training},'' \emph{Proc. VLDB
  Endow.}, vol.~13, no.~12, p. 3005–3018, Aug. 2020. [Online]. Available:
  \url{https://doi.org/10.14778/3415478.3415530}
\BIBentrySTDinterwordspacing

\bibitem{pytorch_blog}
\BIBentryALTinterwordspacing
janeyx99, ``Fsdp \& cudacachingallocator: an outsider newb perspective,'' 2023.
  [Online]. Available:
  \url{https://dev-discuss.pytorch.org/t/fsdp-cudacachingallocator-an-outsider-newb-perspective/1486}
\BIBentrySTDinterwordspacing

\bibitem{pytorch_github_issue}
\BIBentryALTinterwordspacing
A.~Gu, ``{[RFC] Per-Parameter-Sharding FSDP \#114299},'' 2025. [Online].
  Available: \url{https://github.com/pytorch/pytorch/issues/114299}
\BIBentrySTDinterwordspacing

\bibitem{softermax_paper}
\BIBentryALTinterwordspacing
J.~R. Stevens, R.~Venkatesan, S.~Dai, B.~Khailany, and A.~Raghunathan,
  ``{Softermax: Hardware/Software Co-Design of an Efficient Softmax for
  Transformers},'' in \emph{Proceedings of the 58th Annual ACM/IEEE Design
  Automation Conference}, ser. DAC '21.\hskip 1em plus 0.5em minus 0.4em\relax
  IEEE Press, 2022, p. 469–474. [Online]. Available:
  \url{https://doi.org/10.1109/DAC18074.2021.9586134}
\BIBentrySTDinterwordspacing

\bibitem{milakov2018online}
M.~Milakov and N.~Gimelshein, ``{Online Normalizer Calculation for Softmax},''
  \emph{arXiv preprint arXiv:1805.02867}, 2018.

\bibitem{amd_cdna1_2020}
{AMD}, ``{AMD ROCm™ GPU Architecture Overview},'' \url{
  https://rocm.docs.amd.com/en/docs-6.1.1/conceptual/gpu-arch.html }, Apr.
  2024, last updated April 25, 2024; accessed 2025-06-09.

\bibitem{amd_cdna2_2023}
------, ``{AMD CDNA™ 2 Architecture White Paper},'' \url{
  https://www.amd.com/content/dam/amd/en/documents/instinct-business-docs/white-papers/amd-cdna2-white-paper.pdf
  }, Advanced Micro Devices, Inc., Technical Report, 2023, published
  approximately 2.2 years ago (circa early 2023); accessed 2025-06-09.

\bibitem{amd_cdna3_2023}
------, ``{AMD CDNA™ 3 Architecture White Paper},'' \url{
  https://www.amd.com/content/dam/amd/en/documents/instinct-tech-docs/white-papers/amd-cdna-3-white-paper.pdf
  }, 2023, accessed: 2025-06-09.

\bibitem{amd_rocblas_latest}
------, ``{rocBLAS – ROCm Basic Linear Algebra Subprograms Library},''
  \url{https://rocm.docs.amd.com/projects/rocBLAS/en/latest/}, accessed:
  2025-06-09.

\bibitem{amd_matrix_instruction_calculator_2023}
{AMD ROCm Team}, ``{AMD Matrix Instruction Calculator},'' \url{
  https://github.com/ROCm/amd_matrix_instruction_calculator}, 2023.

\bibitem{gqa_paper}
J.~Ainslie, J.~Lee-Thorp, M.~de~Jong, Y.~Zemlyanskiy, F.~Lebron, and
  S.~Sanghai, ``{{GQA}: Training Generalized Multi-Query Transformer Models
  from Multi-Head Checkpoints},'' in \emph{Empirical Methods in Natural
  Language Processing}, 2023.

\bibitem{amd_aig_aima_pytorch_benchmark_2025}
{AMD‑AIG‑AIMA Team}, ``{PyTorch Training Benchmark},'' \url{
  https://github.com/AMD-AIG-AIMA/pytorch-training-benchmark}, 2025.

\bibitem{fsdp2_tutorial}
Y.~M. Wei~Feng, Will~Constable, ``{Getting Started with Fully Sharded Data
  Parallel (FSDP2)},''
  \url{https://docs.pytorch.org/tutorials/intermediate/FSDP\_ tutorial.html},
  2025.

\bibitem{dao2023flashattention}
T.~Dao, ``{Flashattention-2: Faster Attention with Better Parallelism And Work
  Partitioning},'' \emph{arXiv preprint arXiv:2307.08691}, 2023.

\bibitem{amd_hpcfund_hardware_2025}
{AMD HPC Fund Team}, ``{HPC Fund Research Cloud: Hardware Configuration},''
  \url{https://amdresearch.github.io/hpcfund/hardware.html}, 2025.

\bibitem{rocm_rocprofiler_sdk_2025}
{AMD ROCm Team}, ``{ROCprofiler-SDK: Application Profiling, Tracing, and
  Performance Analysis},'' \url{https://github.com/ROCm/rocprofiler-sdk}, 2025.

\bibitem{rocm_roctracer_2025}
------, ``{ROCm ROC‑Tracer: Callback/Activity Library for GPU Performance
  Tracing},'' \url{https://github.com/ROCm/roctracer}, 2025.

\bibitem{rocprofiler-compute}
X.~Lu, C.~Ramos, F.~Zheng, K.~W. Schulz, J.~Santos, K.~Lowery, N.~Curtis, and
  C.~D. Pietrantonio, ``{ROCm/rocprofiler-compute: v3.1.0 (12 February
  2025)},'' \url{https://doi.org/10.5281/zenodo.7314631}, 2025.

\bibitem{semianalysis_mi300x_h100_h200_2024}
{SemiAnalysis}, ``Mi300x vs h100 vs h200 benchmark part 1: Training – cuda
  moat still alive,'' \url{
  https://semianalysis.com/2024/12/22/mi300x-vs-h100-vs-h200-benchmark-part-1-training/#single-node-training-performance
  }, Dec. 2024, published December 22, 2024; accessed 2025-06-13.

\bibitem{amd_rocm_performance_results_2025}
{AMD}, ``{Performance Results with AMD ROCm™ Software},'' \url{
  https://www.amd.com/en/developer/resources/rocm-hub/dev-ai/performance-results.html
  }, May 2025.

\bibitem{intra_iter_launch}
J.~Ekelund, S.~Markidis, and I.~Peng, ``{ Boosting Performance of Iterative
  Applications on GPUs: Kernel Batching with CUDA Graphs },'' in \emph{2025
  33rd Euromicro International Conference on Parallel, Distributed, and
  Network-Based Processing (PDP)}, 2025.

\end{thebibliography}

\end{document}